\documentstyle[pra,aps]{revtex}
\begin{document}
\preprint{\today}
\draft
%
%
\title{Testing the equivalence principle through \\
 freely falling quantum objects}
\author{Lorenza Viola and Roberto Onofrio}
\address{Dipartimento di Fisica ``G. Galilei'', 
Universit\`a di Padova, and INFN, Sezione di Padova, \\ 
Via Marzolo 8, Padova, Italy 35131}
\date{\today}
\maketitle
%
%
\begin{abstract}
Free fall in a uniform gravitational field is revisited
in the case of quantum states with and without classical analogue. 
The interplay between kinematics and dynamics in the evolution of a falling 
quantum test particle is discussed allowing for a better understanding 
of the equivalence principle at the operational level.
\end{abstract}
%
%
%
%
\section{Introduction}
Gravity appears to be distinguished from all the other fundamental 
interactions by the remarkable feature of affecting all bodies 
in a universal way, regardless of their internal composition and mass.  
This fact, which requires in the Newtonian picture 
the gravitational force to be exactly proportional to the inertial mass, 
represents the physical cornerstone of Einstein's weak equivalence principle, 
estabilishing the local identification of gravity and acceleration as 
far as mechanical effects are concerned. 
In the famous Gedankenexperiment conceived by Galileo, 
the universality of the ratio between gravitational 
and inertial masses has been studied by imagining test 
bodies in free fall from the tower of Pisa \cite{GALILEO}. 
Since that time several actual tests of the weak 
equivalence principle have been performed with very sensitive schemes, 
such as the ones exploiting pendula or torsion balances \cite{WILL}, 
but only classical test bodies have been involved.
On the other hand, any attempt to merge quantum mechanics and gravity 
on an operational basis should start from the possibility of estabilishing 
the properties of the latter by using a generic body as a probe, 
regardless of its macroscopicity and hopefully without reference to 
classical physics.
This path of reasoning leads to conceptual difficulties 
clearly focused, among the others, by Roger Penrose when he writes: 
``We see that this view of reality is very different from the one 
that we have become accustomed to from classical physics, where 
particles can be only in one place at a time, where the physics is 
local (except for action at a distance) and where each particle is a 
separate individual object which, when it is in free flight, can be 
considered in isolation from any other particle. All these classical 
conceptions must be overturned once we accept the reality of the 
state-vector. It is perhaps little wonder that most people are reluctant 
to do this" \cite{PENROSE}. 
As a first step along this direction, it is therefore natural to ask
what happens if the Pisa Gedankenexperiment is repeated 
using properly prepared quantum test particles.  
This question, besides its abovementioned conceptual importance, 
is also relevant in view of recent attempts and proposals to investigate 
gravitation using microscopic and mesoscopic systems, such as antimatter 
in free fall \cite{FAIRBANK,NIETO,CERN}, cooled atoms in optical 
molasses \cite{KASEVICH1} and  opto-gravitational 
cavities \cite{AMINOFF,ONVIO}, interfering matter waves  both in 
curved space-time and accelerated reference frames \cite{KASEVICH2,AUDRETSCH}.
The content of the paper is organized as follows. In Section II, 
after some remarks on the preparation of the initial state for a Galileo 
experiment in the classical case, 
a similar prescription is discussed for all the quantum states 
mapped, in the macroscopic limit, into classical states.  
In Section III the peculiar case of quantum states without 
a classical analogue is dealt, with a detailed analysis of the simplest 
class of Schr\"odinger cat states in the configurational space. 
In Section IV the two lowest momenta of the time of flight distributions 
for quantum states in free fall in a uniform gravitational field are 
evaluated. The dynamical effect of a continuous quantum measurement of 
position on the geodesic motion is dealt in Section V. 
In Section VI some phenomenological consequences of the previous 
considerations are presented, with particular emphasis on 
the possibility of testing gravity with mesoscopic objects.  
Some general comments on the definitions of the weak equivalence 
principle which still hold in the quantum case are finally discussed. 
 
\section{The Galileian preparation for classical-like states}

In order to perform an ideal free fall experiment for two quantum 
particles having inertial masses $m_i^{(1)}$ and $m_i^{(2)}$, 
$m_i^{(1)} \neq m_i^{(2)}$, we have first of all to specify a proper 
initial preparation in such a way that any difference in the motion 
during the free fall must be ascribed to the effect of gravity. 
By recalling that within the classical Hamilton 
picture the Galileian prescription for initial positions and velocities 
fixes the ratio between the initial momenta in a well defined way, 
$p_0^{(1)}/p_0^{(2)}= m_i^{(1)}/m_i^{(2)}$, it is natural to extend such 
a prescription to the quantum case, which can be also represented 
through a Hamiltonian scheme. 
Of course, the Heisenberg uncertainty principle prevents to simultaneously 
define, for each particle, initial position and momentum. 
If $|\psi_1 \rangle$ and $|\psi_2 \rangle$ 
denote the initial state vectors for particles 1, 2 in the Schr\"odinger 
picture, the classical recipe can be reasonably rephrased by imposing 
the following conditions:
\begin{equation}
\langle \hat{z} \rangle_{\psi_1} = \langle \hat{z} \rangle_{\psi_2} \:,
\hspace{1cm}
{\langle \hat{p}_z \rangle_{\psi_1} \over m_i^{(1)} } =
{\langle \hat{p}_z \rangle_{\psi_2} \over m_i^{(2)} }  \:, 
\label{GALILEO}
\end{equation}
where for simplicity we have  restricted ourselves to a one-dimensional 
representation along the vertical $z$-direction and 
$\langle \hat{z} \rangle_{\psi}$, $\langle \hat{p}_z \rangle_{\psi}$
denote, as usual, the expectation values for position and momentum 
operators respectively. Furthermore, our description will be confined to 
the motion of nonrelativistic quantum particles, implying values of the 
initial velocities in (\ref{GALILEO}) small compared to the speed of light.
From the mathematical viewpoint, equation 
(\ref{GALILEO}) imposes a relative constraint on the average values of the 
position and momentum probability distributions associated to the states 
$|\psi_1 \rangle$ and $|\psi_2 \rangle$ respectively. 
Some other remarks are in order. 
First, leaving aside the special choice of starting from a 
given position at rest ($\langle \hat{p}_z \rangle_{\psi_1} = 
\langle \hat{p}_z \rangle_{\psi_2}=0 $), we are forced in general to deal with two different 
initial states $|\psi_1 \rangle$ and $|\psi_2 \rangle$ in the single particle Hilbert 
space; this is analogous to the classical situation, where 
the representative point of the  classical initial state in the 
phase space is different for the two masses. Second, 
deep differences between the concepts of classical and quantum state 
exist, which also produce differences in the corresponding prepared systems. 
The probabilistic interpretation underlying quantum mechanics 
will allow to speak of probability distributions for instance 
characterized by {\sl mean} initial conditions 
like (\ref{GALILEO}), as opposite to the well defined values 
for the relevant classical observables. 
Moreover, as a striking difference 
with respect to the classical case, conditions (\ref{GALILEO}) are far from 
univocally determining the initial state of the two particles. On the contrary, 
the Galileian prescription gives rise to equivalence classes of states in the 
Hilbert space, each class possessing a defined mean position and velocity and 
collapsing  in the classical limit into a state having both quantities 
sharply defined. It is also rewarding to
point out that this line of reasoning, being ultimately motivated by the 
correspondence principle, applies in a strict sense to all quantum states 
for which a classical interpretation is obtainable. 
Such a class of states does not exhaust the totality of the admissible ones 
in the Hilbert space. For all the other states, namely the family of 
genuinely quantum states for which no classical counterpart exists and the 
correspondence principle cannot be invoked, the Galileian prescription 
(\ref{GALILEO}) has to be {\sl postulated} and deserves therefore a more 
careful discussion. 

\section{The case of states having no classical analogue}

Due to the superposition principle, states of intrinsically 
quantum nature arise even starting from states which in the classical 
limit corresponds to macroscopically distinguishable ones.  
Let $|\psi_n \rangle$, $n=1,\ldots, N$, denote a set of states in the 
Hilbert space of a given quantum system, and let us suppose these states 
to be macroscopically distinct. Any superposition state 
$|\psi_0 \rangle = \sum_n c_n |\psi_n \rangle$ is also permitted, the 
complex coefficients $c_n$ ensuring the overall normalization.
The non-classical nature of this superposition state can be made 
explicit by considering the density matrix representation, 
\begin{equation}
\hat{\rho}= |\psi_0 \rangle \langle \psi_0| =
\sum_n |c_n|^2 |\psi_n \rangle \langle \psi_n| +
\sum_{n,m \not = n} c_n^{\ast} c_m |\psi_n \rangle \langle \psi_m| \:,  
\label{RHO}
\end{equation}
the off-diagonal terms being responsible for correlations of a purely 
quantum mechanical origin. 
In the classical limit, due to the action of decoherence mechanisms, 
interference effects are lost and the pure state description (\ref{RHO}) 
becomes identical to the statistical mixture characterized by the 
diagonal probability weights $|c_n|^2$ alone. Given the superposition 
state $|\psi_0 \rangle$, the separation indicated in (\ref{RHO}) between 
diagonal and off-diagonal contributions also reflects on observable 
properties of the system, such as average values of Hermitian operators. 
For a generic observable $\hat{\cal O}$, one can think its overall mean 
value in the state $|\psi_0\rangle$ as formed, according to (\ref{RHO}), 
by two distinct terms:
\begin{equation}
\langle \hat{\cal O} \rangle_{\psi_0} = 
\langle \hat{\cal O} \rangle_{\mbox{\small classical}} +
\langle \hat{\cal O} \rangle_{\mbox{\small purely quantum}}  \:.
\label{DECOMP}
\end{equation} 
If only non-diagonal entries are different from zero in (\ref{DECOMP}), 
($\langle \psi_n | \hat{\cal O} | \psi_n \rangle =0$ for all $n$), 
the mean value of the observable $\hat{\cal O}$ has only contributions 
of intrinsically quantum mechanical origin.

Coming back to the free fall problem, the previous considerations apply 
if a non-classical superposition state is selected as initial state for 
one or both test particles $m^{(1)}_i$, $m^{(2)}_i$, the decomposition 
(\ref{DECOMP}) holding in this case for the position and momentum 
operators $\hat{z}$, $\hat{p}_z$ involved in (\ref{GALILEO}). Despite of 
the fact that purely quantum expectation values may emerge, 
compatibility with the classical limit is again maintained provided that the  
Galileian conditions (\ref{GALILEO}) are satisfied, in the sense that both 
states are mapped into initial configurations having the same (classical) 
position and the same (classical) velocity. 

As is well known, a nice class of quantum states without classical counterpart 
is offered by the so-called Schr\"odinger cat states, firstly 
introduced in \cite{SCHRO}. 
Let us consider the coherent superposition of 
two macroscopically distinguishable states in the configurational space, 
represented by a wavefunction of the following form:
\begin{equation}
\psi_0(z)=N \bigg\{ c_+ \exp\bigg( -{(z-z_0+\Delta)^2 \over 2 \Delta_0^2} 
\bigg)  
+ c_- \exp\bigg( -{(z-z_0-\Delta)^2 \over 2 \Delta_0^2} \bigg) \bigg\}\:, 
\label{CAT}  
\end{equation}
consisting of the sum of two Gaussian peaks at $z=z_0 \pm \Delta$, 
$\Delta >0$, each of width $\Delta_0$. Here $c_{\pm}$ are two complex 
coefficients, for which we denote by $\vartheta$ the relative phase, 
and $N$ is a normalization constant determined (up to an irrelevant phase
factor) by
\begin{equation}
|N|^2 =(\pi \Delta_0^2)^{-1/2} \big \{ |c_+|^2 + |c_-|^2 +2 \mbox{Re}(c_+ 
c_-^*)\exp(-\Delta^2/ \Delta_0^2) \big\}^{-1} \:.
\end{equation}
For null separation ($\Delta=0$) the special case of a normalized Gaussian 
wavepacket with vanishing average momentum is recovered. The 
expectation values of position and momentum in the state (\ref{CAT}) are 
calculated obtaining:
\begin{equation}
\langle \hat{z} \rangle_{\psi_0}= 
z_0 -\Delta { |c_+|^2-|c_-|^2  \over
{|c_+|^2+|c_-|^2+2 \mbox{Re}(c_+c_-^*)\exp(-\Delta^2/ \Delta_0^2)}} \:,
\label{AVEQ}
\end{equation}
\begin{equation}
\langle \hat{p}_z \rangle_{\psi_0}= 
-2 \hbar {\Delta \over \Delta_0^2} 
{{\mbox{Im}(c_+ c_-^*)\exp(-\Delta^2/ \Delta_0^2)} \over
{|c_+|^2+|c_-|^2+2 \mbox{Re}(c_+c_-^*)\exp(-\Delta^2/ \Delta_0^2)}} \:.
\label{AVEP}
\end{equation}
The average position is simply the center of mass of the cat, weighted 
by the asymmetry between the coefficients $|c_+|^2$ and $|c_-|^2$. 
By Fourier transforming (\ref{CAT}), it is seen that
no diagonal momentum contributions are present, leading to a purely quantum 
momentum in (\ref{AVEP}) with the form of a typical interference factor,
$ \mbox{Im}(c_+ c^*_-) = |c_+| |c_-| \sin \vartheta$. A vanishing value 
of $\langle \hat{p}_z \rangle_{\psi_0}$ is found if the relative phase 
$\vartheta = k \pi$, $k \in Z$, corresponding to the even (male) and 
odd (female) combinations of definite parity cat states \cite{MANKO}; 
a maximum contribution is instead achieved at $\vartheta= (2 k+1)\pi/2$, 
$k \in Z$, corresponding to cat wavefunctions already introduced by Yurke 
and Stoler \cite{YURKE}. In order to study the free fall motion of two 
cat-like quantum particles, one has to match prescriptions (\ref{GALILEO}) 
 by suitably gauging 
the parameters of the wavefunction (\ref{CAT}) which is, at least in 
principle, possible. Let us  suppose for definiteness to maintain 
the width $\Delta_0$ fixed and $|c_+|=|c_-|$, in such a way that the 
initial average position (\ref{AVEQ}) is constrained to be $z_0$. 
The simplest choice is clearly represented by parity eigenstates, i.e. 
by even/odd cat states necessarily starting at rest. If, however, 
non-vanishing average momenta are present, one can exploit the possibility 
to tune the separation $\Delta_{1,2}$ between the peaks, with 
$\vartheta$ fixed, i.e. 
\begin{equation}
{\Delta_2 \over \Delta_1} \exp \bigg( - \frac{\Delta^2_2- 
\Delta^2_1}{\Delta_0^2} \bigg)
{ {1+\cos \vartheta \exp{(-\Delta_1^2/\Delta_0^2)} } \over 
  {1+\cos \vartheta \exp{(-\Delta_2^2/\Delta_0^2)} } }
= {m_i^{(2)} \over m_i^{(1)} }  \:, 
\end{equation}
or the relative quantum phase $\vartheta_{1,2}$, with $\Delta$ fixed, i.e.
\begin{equation}
{{\sin \vartheta_2 (1+ \exp {(-\Delta^2/\Delta_0^2)} \cos\vartheta_2)}\over 
 {\sin \vartheta_1 (1+ \exp {(-\Delta^2/\Delta_0^2)} \cos\vartheta_1)}}
= {m_i^{(2)} \over m_i^{(1)}} \:.
\end{equation}

Having assigned the initial preparation of each particle, we will 
analyze in the following two sections the evolution of the system during 
the free fall as  respectively predicted by ordinary quantum mechanics 
for closed systems and by a more general description also 
including the effect of a measuring apparatus.

\section{Running the gedankenexperiment: the unmeasured evolution}
 
In the nonrelativistic picture we are considering, the time 
evolution is generated by the Hamiltonian, 
$\hat{H}=\hat{p}_z^2/2 m_i  + m_g g \hat{z}$, 
$m_g$ denoting the gravitational mass. In analogy with Galileo's 
procedure let us focus, for instance, on the time of flight of the 
quantum particles.
Due to the already remarked fuzziness of quantum 
states, even if the measurements are treated as ideal, 
allowing to estimate all the measured quantities with infinite 
precision, these times can be only predicted in a probabilistic way.
Given our particles with inertial masses $m_i^{(1)}$ and $m_i^{(2)}$, 
the experiment consists in recording two time of flight 
distributions corresponding to the time of arrival from the initial 
height $z_0={\langle \hat{z} \rangle}_{\psi_0}$, 
at variance with the classical case where two single time 
values are sampled. Therefore, the full knowledge of such time of flight 
distributions can in principle be demanded to gain a complete 
information on the free fall dynamics.
To the best of our knowledge, no general consensus has been reached 
so far on a consistent definition of time of flight probability density. 
In the probabilistic language, the question is related to the so-called first 
passage time problem, which has been solved within the natural framework 
offered by stochastic mechanics \cite{NELSON} in the special case of 
stationary states \cite{TRUMAN}. 
Various attempts in different directions have been made, 
as indicated in a recent work on the subject where an operatorial solution 
is proposed \cite{ROVELLI1}. Leaving aside a rigorous derivation 
which is not essential to sketch our main line of reasoning, we will 
now limit ourselves to simple arguments. 
The average time of flight for the test mass $m_i^{(k)}$ can by 
straightforwardly calculated by means of the Ehrenfest theorem which, 
owing to the linearity of the gravitational potential, allows to obtain 
the average position at time $t$ in the classical form :
\begin{equation}
\langle z^{(k)}(t) \rangle = -{1 \over 2} {m_g^{(k)} \over m_i^{(k)} } g t^2 
+ {\langle \hat{p}_z \rangle_{\psi_0} \over m_i^{(k)}} t + \langle \hat{z} 
\rangle_{\psi_0} \:, \hspace{1cm} k=1,2\:.    
\label{FREEFALL}
\end{equation}
By setting $\langle z^{(k)}(t) \rangle=0$, the mean time of flight at 
ground level is obtained; in the particular case 
$\langle \hat{p}_z \rangle_{\psi_0}=0$, the corresponding expression is 
\begin{equation}
T_{of}^{(k)} = \sqrt{ 2\bigg( {m_i^{(k)} \over m_g^{(k)}} \bigg) 
{z_0 \over g} } \:, \hspace{1cm} k=1,2 \:.
\label{TOF}
\end{equation}
A rough estimate of the fluctuations around this mean value, taking into 
account the spreading of the state during the motion, can be given by 
evaluating  $\sigma_{T_{of}} \approx \sigma_z(T_{of})/v_z(T_{of})$, where 
$\sigma^2_z(T_{of})$ and $v_z(T_{of})$ are the position variance of the 
state and the average velocity at time $t=T_{of}$ respectively. 
By exploiting (\ref{FREEFALL}), this may 
be shown to be equivalent to another possible definition of 
$\sigma_{T_{of}}$, resulting from the semidifference between the times 
$t_+$, $t_-$ for which $\langle z(t_{\pm}) \rangle \pm \sigma_z(t_{\pm}) =0$. 
Let us consider in detail the behavior of our 
quantum probes $m_i^{(1)}$ and $m_i^{(2)}$, when each of them is allowed 
to be initially prepared with average position $z_0$ and vanishing 
average momentum in the form of a Gaussian or a cat-like state (this 
latter necessarily possessing a well defined parity). The general 
expression of the position variance as a function of time, 
$\sigma^2_z(t)= \langle \hat{z}^2 \rangle_{\psi_t} -
\langle \hat{z} \rangle_{\psi_t}^2$, has been calculated by 
following the Schr\"odinger evolution having the wavefunction 
(\ref{CAT}) as initial condition (see Appendix A for details). The result 
of the calculation can be written as follows:
\begin{equation}
\sigma_z(t) = \sqrt{ {\Delta_0^2 \over 2} + 
{ \Delta^2 \over 1\pm \mbox{e}^{-\Delta^2/\Delta_0^2} } + 
{ \hbar^2 \over 2 m_i^2 \Delta_0^2 } \bigg( 1 \mp 2 
{\Delta^2 \over \Delta_0^2 } { \mbox{e}^{-\Delta^2/\Delta_0^2} \over 
1\pm \mbox{e}^{-\Delta^2/\Delta_0^2} } \bigg) t^2 } \:,
\label{VARIANCE}
\end{equation} 
the upper and lower signs referring to the even (male, $c_+$=$c_-$=1) and 
odd (female, $c_+$=$- c_-$=1) cat state respectively. 
As we will 
discuss in more detail elsewhere \cite{ONVIO2}, the rate of 
spreading for a cat state may become less than the Gaussian one. 
Eq. (\ref{VARIANCE}) reduces to the familiar formula 
for the spreading of Gaussian wavepackets if $\Delta=0$, while 
slight modifications to the Gaussian case are obtained in 
the limit $\Delta/\Delta_0 \gg 1$. Let us then consider an 
intermediate regime with $\Delta \approx \Delta_0$, moreover assuming 
$\Delta_0$ small enough to become 
negligible with respect to the evolution induced contribution in the 
total spreading $\sigma_z(t)$. One can thus approximate Eq. 
(\ref{VARIANCE}) by the leading time-dependent term and collect the 
available information concerning the time of flight distribution in the 
form of average time and standard deviation as:
\begin{equation}
T_{of} \pm \sigma_{T_{of}} = \sqrt{ 2 
\bigg( {m_i^{(k)} \over m_g^{(k)}} \bigg) {z_0 \over g} } \pm {\sqrt{2}\over 2}
\epsilon {\hbar \over \Delta_0 m^{(k)}_g g }\:, \hspace{1cm} k=1,2\:,
\label{DISTRIB}
\end{equation}
where $k=1,2$ distinguishes, as usual, the two particles and $\epsilon$ 
is a numerical factor given by
\begin{equation}
\epsilon = \left\{ \begin{array}{cl}
                   1  &  \mbox{Gaussian state,} \\
\left( {\mbox{e}-1 \over \mbox{e}+1 } \right)^{\pm 1/2} &
                         \mbox{Male ($+$)/Female ($-$) cat state.}
                   \end{array} \right. 
\end{equation}
Eq. (\ref{DISTRIB}) can in fact be used with a general validity, 
provided the final time in (\ref{VARIANCE}) is long enough, 
which can in turn be obtained by properly adjusting the initial height. 
It is manifest from (\ref{DISTRIB}) that, despite the semiclassical 
preparation recipe (\ref{GALILEO}), the time of flight distributions 
corresponding to different quantum objects are different, 
due either to different masses or different initial states. As an 
interesting feature, the ratio between inertial and 
gravitational mass contributes to the average time of flight, whereas 
the fluctuations around this value are affected, at this stage, by the 
{\sl gravitational} mass alone. 

Since Eq. (\ref{TOF}) holds, a necessary and sufficient condition in order 
to conclude
\begin{equation}
{m_i^{(1)} \over m_g^{(1)}} ={m_i^{(2)} \over m_g^{(2)}} = \mbox{
constant} 
\label{RATIO}
\end{equation}
is to observe, in close analogy with the classical experiment, identical 
values for the average times of flight. Let us suppose that the equality 
(\ref{RATIO}) has been established for two particles initially prepared 
in the same type of state, for instance a Gaussian one. 
At first sight one may wonder that, at variance with the classical behavior, 
a dependence upon the mass survives in time of flight spreading, 
allowing one to distinguish again the resulting distribution patterns. 
At a closer inspection, the role and nature of the mass dependence can  
be understood on the basis of kinematical arguments. Whatever such a 
dependence will be, the crucial point is that it can be recovered from 
an equivalent problem where it has a completely kinematical origin. 
Let us imagine a laboratory in which a test particle of mass 
$m_i$ is initially placed on the top, at a height $z_0$ above the floor, 
with the laboratory being then propelled at uniform acceleration 
$\vec{a}$ with respect to a inertial reference frame, $\vec{a}$ directed upward.
The motion of the particle, which is supposed to freely evolve in 
the non inertial frame, can be derived from the inertial free motion by 
a standard procedure, as outlined in Appendix B. Provided that 
equality (\ref{RATIO}) is fulfilled and $a=g$, formally identical 
evolution equations are thus found either for the motion of a quantum particle 
subjected to a uniform gravitational field with strength $g$ in a 
inertial reference, or for the motion of the same particle freely evolving 
in a non inertial reference accelerated with $a=g$.
In particular, an observer inside the accelerating 
laboratory should see the particle hitting the ground level after the same 
average time as in (\ref{DISTRIB}) but with a variance proportional 
to the inertial mass, the only property of the body available in this case. 
More generally, once Eq. (\ref{RATIO}) is fulfilled, the theory shows a 
complete identification between the effects of gravitation and 
acceleration, predicting in particular that time of flight 
probability distributions of identical form, with identical mass 
dependence in {\sl every} momentum, not just in the second one as here 
considered, are detected if the motion is performed in the gravitational 
or accelerated laboratory.   
  
By summarizing, the widespread quoted sentence according to which all bodies 
equivalently prepared fall precisely the same way in a gravitational field 
has to be carefully interpreted when quantum objects are considered. 
Unlike the classical case, this does not imply that only mass independent 
observables are found. On the contrary, the time of flight probability 
distributions of quantum particles remain mass-dependent, but such a 
dependence is  {\sl expected in order the weak equivalence principle 
be preserved at the quantum level}. 

\section{The measured evolution}

Until now, the discussion has been carried out without considering the 
effect of the measurement apparatus. Indeed, detection schemes can be 
designed in which each falling atom impulsively interacts with the meter 
just before stopping its evolution, that is at the arrival time.
However, experimental situations involving a continuous 
monitoring of the atom throughout the whole free fall can be investigated (see 
\cite{BORDE} for recent proposals) and the perturbation introduced by the meter 
cannot be reduced without that a parallel limitation on the extracted 
information also results. 
Such an influence has then in principle to be taken into account, as we are 
going to discuss in this section. Various models have been designed to include 
the effect of a continuous measurement process into the dynamics, the 
so-called measurement quantum mechanics, an account of which can be found 
for instance in \cite{MQM}. In the so-called non-selective approach, i.e. 
when no particular history of measurement is selected, the evolution of the 
system under a continuous measurement of position $\hat{z}$ is described 
through a master equation for the reduced density matrix operator:
\begin{equation}
{d \over dt} \hat{\rho}(t)= - {i \over \hbar} 
[ \hat{H}, \hat{\rho}]  -
{\kappa_z \over 2}  [\hat{z},[\hat{z},\hat{\rho}]] \:, \label{MASTER}
\end{equation}  
being $\hat{H}$ the Hamiltonian of the system, $[\,,\,]$ 
the commutator, and $\kappa_z$ expressing (in $(m^2/\mbox{Hz})^{-1})$ the 
coupling of the position meter to the test particle. 
Hereafter, $\kappa_z$ is assumed to be time-dependent.
Eq. (\ref{MASTER}) for the system under consideration can be rewritten 
as:
\begin{equation}
{\partial \over \partial t} \rho(z,z',t)= \bigg\{ {i \hbar \over 2m} 
\bigg( {\partial^2 \over \partial z^2} - 
{\partial^2 \over \partial {z'}^2}\bigg) - {i m_g g \over \hbar} (z-z') 
- {\kappa_z \over 2}(z-z')^2 \bigg\} \rho(z,z',t) \:,  \label{MASTER2}
\end{equation}  
with $\rho(z,z',t)=\langle z|\hat{\rho}(t) |z'\rangle$ denoting the 
coordinate representation of the density operator. 
Besides some numerical factor, the coupling constant $\kappa_z$ is equal 
to  the reciprocal of the position noise spectral density. If this last 
is decreased, i.e. the position sensitivity is increased, the last 
term in the righthandside of (\ref{MASTER2}) will dominate over the 
others. As remarked above, under this assumption the considerations 
made in the previous section can be affected by the measurement 
process, and have to be reanalyzed in detail.  A straightforward but 
tedious calculation (see Appendix A for more details) allows one to 
explicitely solve the master equation (\ref{MASTER2}) when the initial 
wavefunction has a Gaussian or cat-like form. In particular, it is 
possible to calculate the average position and its variance versus 
time. It turns out that, while the former is left 
unchanged with respect to the unmeasured case (\ref{FREEFALL}), the 
position variance is modified by the measurement coupling, the net 
effect being represented by an additive time dependent contribution of 
the following form:
\begin{equation}
\sigma^2_z(t;\kappa_z)=\sigma^2_z(t;\kappa_z=0) + {\kappa_z \over 3} 
{\bigg( {\hbar \over m_i} \bigg)}^2 t^3  \:, 
\end{equation}
with $\sigma^2_z(t;\kappa_z=0)$ given in (\ref{VARIANCE}). It is worth to 
observe that an additional mass dependence is obtained in this case.
Despite of the fact that such a dependence is not different  
from the unmeasured one, the ratio $\hbar/m_i$ still appearing, an 
explanation analogous to the one delined at the end of the previous 
section requires care, due to the physical meaning of the coupling 
$\kappa_z$. In order to state the problem in an accelerated frame some 
additional assumptions on the behavior of such a parameter are required. 
Since in general the coupling between the meter and the test mass is not a 
purely mechanical one, the question can be properly addressed only 
by {\sl postulating} the validity of the strong equivalence principle. 

The previously noted lack of effect on the average position is 
interpreted as a manifestation of the Ehrenfest theorem for potentials 
written as polinomials up to the third power of distance \cite{CAVES}. 
However, for motions in more complicated gravitational fields, 
the last term in the righthandside of (\ref{MASTER2}) can create 
differences in the average position of particles having different 
masses, something like a {\sl gravitational} quantum Zeno effect.
This would be on the other hand the 
signal of a contrast between a body at the same time being in free
fall and having its position continuously registered via 
a meter interacting with it. 
Tests of the equivalence principle, although still viable, become more 
complicated since disentanglement of the effect of the meter from the 
purely gravitational one is required.
This problem is also present in the case of other tests 
of the equivalence principle such the ones exploiting rigid objects. 
For such configurations additional complications arise from the
difficulty to achieve the quantum domain in macroscopic bodies \cite{BOCKO}.

\section{Phenomenology}

Recent progress in the manipulation of atomic states gives hope to make 
the considerations presented here less remote from experimental investigation 
than expected. Atomic mixed states with a Gaussian phase space 
distribution, although far from the minimum uncertainty value of the 
pure state configuration, have been already prepared and used to study the 
free fall of atoms in a semiclassical regime, as reported in \cite{AMINOFF}. 
A cloud of Cesium atoms was trapped, cooled at a temperature of a few 
$\mu$K and then released at an average height of 2.91 mm above an atomic 
mirror made by a dielectric surface and a repulsive evanescent field. 
Various bounces of the cloud were observed, and the time of flight was 
measured. In a successive experiment, a vibrating mirror was used to show 
phase modulation of atomic waves. In this case, an accurate study of the 
time of flight was reported, showing that resolutions of the order of 
0.5 ms  can be achieved \cite{DALIBARD}. 
On the other hand, preparation of even and odd superposition states has been 
proposed  \cite{VOGEL}, and Schr\"odinger cats of a single trapped 
$\mbox{Be}^+$ ion with a separation $\Delta \approx 10^2$ nm have been 
recently generated and detected in laboratory \cite{WINELAND}. 
Merging these accomplishments, an experiment in which a Gaussian or a 
Schr\"odinger cat state of matter at the single atom level is bouncing over 
an atomic mirror can be envisaged. 
Some numerical values, evaluated from (\ref{DISTRIB}) for different 
atomic species routinely manipulated in laboratory, are reported in 
Table I, allowing to clarify the orders of magnitude involved in a 
possible experimental test. 
The already achieved time of flight resolution quoted in \cite{DALIBARD} 
allows one to observe different time of flight distributions either due to 
different masses (within the same column) or to different states (within the 
same row). Experiments of this kind should also stimulate further thoughts on 
consistent definitions of the time of flight distributions 
in the quantum domain.   
Moreover, similar experiments should be performed by actually observing the 
time of flight distributions in accelerated frames, for instance 
exploiting centrifugal force fields, to verify if the weak equivalence 
principle, as supposed in our discussion, still holds in the 
quantum realm. In performing such experiments, one should be prepared to 
possible surprises, since no evidence is so far available that Nature 
preserves the equivalence principle at the quantum level.

\section{Discussions and conclusions}

Besides the experimental feasibility, some conceptual observations 
concerning the interplay between quantum mechanics and gravitation are 
in order. First of all, the presence of the mass in the time of flight 
distributions cannot be ascribed to the fact that the 
test objects possess, in some sense, an extended structure. 
It is indeed apparent from (\ref{VARIANCE})-(\ref{DISTRIB}) that, due to the 
uncertainty principle, a further increase in the 
spreading is found if more point-like structures (i.e. smaller $\Delta_0$ 
values) are allowed. Such a dependence can be instead deeply related 
to the impossibility of reproducing, for any quantum 
object, the classical concept of {\sl deterministic} trajectory. 
It may be helpful to recall that in Nelson's picture of quantum 
mechanics \cite{NELSON} the kinematics is modelized in terms of 
{\sl stochastic} trajectories in the configuration space. 
Within this framework, it 
is not surprising that the combination $\hbar/m$ ultimately appears in 
Eq. (\ref{DISTRIB}), which is nothing but the Brownian diffusion 
coefficient accounting for the degree of stochasticity of quantum 
kinematics as opposite to the deterministic classical one. As a related 
question, which is preexisting to the introduction of the 
gravitational field itself, this unavoidable stochasticity introduces 
troubles if the same procedure of classical relativity is adopted to 
operatively define geodesics curves and associated inertial frames. In 
the quantum case, the possibility of a simple identification between the 
world lines of freely falling bodies and a set of preferred entities 
with a purely geometric nature clearly no more holds. We refer to 
\cite{ROVELLI2,TOLLER} for a detailed account on the definition of 
reference frames by means of material quantum objects as a preliminary step 
toward quantum gravity; see also \cite{VIOLA} for an attempt to give a 
variational definition of a quantum geodesic within a relativistic 
stochastic scheme. 

In summary, we have discussed a revival of the Galileo free fall 
experiment using quantum test objects. Both the initial preparation 
and the dynamical evolution have been analyzed with special care to 
states of intrinsically quantum nature. It turns out that, despite 
the possibility of {\sl weighting} different quantum objects by looking 
at their free fall evolution, a complete identification between the 
effects of gravitation and acceleration is expected in agreement 
with the equivalence principle. Some troubles may instead emerge 
by including a continuous measurement process, which demands for 
a reformulation of the concept of {\sl free} fall itself.
It is not unlikely that an operative definition of the equivalence 
principle consistent with quantum measurement theory will require the 
emergence of new concepts in gravitation \cite{PENROSE}.  


\acknowledgments 
We are grateful to V. I. Man'ko for bringing reference \cite{VOGEL} 
to our attention. It is also a pleasure to thank C. Rovelli, D. Zanello 
and S. Rufolo for stimulating discussions, F. De Felice and M. Tonin for a 
critical reading of the manuscript. 
This work was supported by INFN, Italy, through the 
Iniziativa Specifica NA12 on Quantum Vacuum and Gravitation.

\appendix

\section{Solution of the master equation with measurement coupling}

The master equation (\ref{MASTER2}) can be conveniently rewritten in terms 
of new independent variables $u$ and $v$ defined as
\begin{eqnarray}
u & = & {z+z' \over 2} \:, \nonumber \\
v & = & z-z' \:, 
\end{eqnarray}
obtaining
\begin{equation}
{\partial \over \partial t} \rho(u,v,t)= \bigg\{ 2i \mu 
{\partial^2 \over \partial u \partial v} - i \nu v 
- {\kappa_z \over 2} v^2 \bigg\} \rho(u,v,t) \:,  \label{MASTER3}
\end{equation}  
being $\mu=\hbar/2 m_i$, $\nu=m_g g/\hbar$. By performing the Fourier 
transform with respect to the variable $u$,
\begin{equation}
\tilde{\rho}(\alpha, v,t)= {1 \over \sqrt{2 \pi}} 
\int_{-\infty}^{+\infty} \mbox{e}^{-i\alpha u} \rho(u,v,t) \,du \:, 
\end{equation}
we get from (\ref{MASTER3}) the equation
\begin{equation}
{\partial \over \partial t} \tilde{\rho}(\alpha,v,t)= \bigg\{ -2 \mu \alpha  
{\partial \over \partial v} - i \nu v - {\kappa_z \over 2} v^2 \bigg \} 
\tilde{\rho}(\alpha,v,t) \:.  \label{MASTER4}
\end{equation}  
The solution of (\ref{MASTER4}) is 
\begin{equation}
\tilde{\rho}(\alpha,v,t)=\tilde{\rho}_0(\alpha, v-2\mu \alpha t)\,
\exp\bigg\{ i \nu(\mu \alpha t^2 -v t) - \kappa_z v^2 t + 2 \mu \alpha 
\kappa_z v t^2 - {4 \over 3} \kappa_z \mu^2 \alpha^2 t^3 \bigg\} \:,
\label{SOLUTION}
\end{equation}
$\tilde{\rho}_0(\alpha,v)$ denoting the Fourier transform of the initial 
density matrix $\rho_0(u,v)$. The average position and the average 
square position at time $t$ are respectively given by 
\begin{equation}
\langle \hat{z}(t) \rangle= \int dz \, z \rho(z,z,t) \:, \hspace{1cm}
\langle \hat{z}^2(t) \rangle= \int dz \, z^2 \rho(z,z,t) \:.
\end{equation}
In case the initial density matrix $\rho_0(z,z')=\psi_0(z)\psi_0(z')^*$ 
arises from a Gaussian or a cat-like wavefunction $\psi_0$ as considered 
in Section III, the Fourier transform (\ref{SOLUTION}) can be explicitely 
inverted and the above quantities analitically evaluated. 
The Ehrenfest expression (\ref{FREEFALL}) for the average position is 
then recovered, whereas the complete expression for the position variance is 
the following:
\begin{eqnarray}
\sigma^2_z(t) &=& 2 A(t) +
 {\Delta^2 \over B} \Big[ |c_+|^2 + |c_-|^2 -
 {1 \over B}( |c_+|^2 - |c_-|^2 )^2 \Big] -   
 {8 \hbar \over B^2 m_i} \Big[ (|c_+|^2 - |c_-|^2) \mbox{Im}(c_+ c_-^*) 
 {\Delta^2 \over \Delta^2_0} \mbox{e}^{-\Delta^2/\Delta_0^2} \Big] t 
  \nonumber \\
& - & {1 \over B} \Big[ {2 \hbar^2 \over m_i^2} \mbox{Re}(c_+ c_-^*) 
  {\Delta^2 \over \Delta^4_0} \mbox{e}^{-\Delta^2/\Delta_0^2} +
  {4 \hbar^2 \over B m_i^2} \mbox{Im}^2 (c_+ c_-^*) 
  {\Delta^2 \over \Delta^4_0} \mbox{e}^{- 2 \Delta^2/\Delta_0^2} \Big] t^2 
\:, \end{eqnarray}
having denoted 
$A(t)=\Delta_0^2/4 + \hbar^2/(4 m_i^2 \Delta^2_0) t^2 + 
\hbar^2/(6 m_i^2) \kappa_z t^3$ and $B=|c_+|^2 + |c_-|^2 + 
2 \mbox{Re}(c_+ c_-^*)\exp{ (-\Delta^2/\Delta_0^2)}$ respectively. 
The unmeasured evolution is obtained when $\kappa_z=0$, while the evolution of 
a Gaussian state corresponds to the choice $\Delta=0$. 

\section{Nonrelativistic Schr\"odinger evolution in accelerated frames}

Let us consider a free nonrelativistic quantum particle, satisfying the 
one dimensional Schr\"odinger equation
\begin{equation}
i \hbar {\partial \psi \over \partial t }(z,t) = - {\hbar^2 \over 2m} 
{\partial^2 \psi \over \partial z^2} (z,t) \:, 
\label{SCHROD}
\end{equation}
and let us introduce an accelerate frame of reference by means of the 
following coordinate transformation:
\begin{eqnarray}
 & & \left\{ \begin{array}{l}
      z'= z - vt - {1 \over 2} a t^2 \:, \\
      t'= t \:,
     \end{array} \right.
\label{TRANSFORM}
\end{eqnarray}
being $v$ and $a$ constant. As a consequence of (\ref{TRANSFORM}), a 
corresponding transformation on the space of states will be induced, 
mapping the wavefunction $\psi(z,t) \rightarrow \psi'(z',t')$. We 
represent such a transformation via the Ansatz
\begin{equation}
\psi'(z',t')=\mbox{e}^{i f(z',t')} \psi(z(z',t'),t(z',t')) \:, 
\label{ANSATZ}
\end{equation}
where the real function $f(z',t')$ has been introduced to allow the 
possibility of a local phase factor and $(z(z',t'),t(z',t'))$ denotes 
the inverse transformation, obtainable from (\ref{TRANSFORM}) by letting 
$v \mapsto -v$, $a \mapsto -a$. The equation of motion satisfied by 
the transformed wavefunction ({\ref{ANSATZ}) with a  generic $f$ can be 
straightforwardly obtained from (\ref{SCHROD}):
\begin{equation}
i \hbar {\partial \psi' \over \partial t'} - \hbar 
{\partial f \over \partial t'} \psi' + (v+a t')
\Big( i \hbar {\partial \psi' \over \partial x'} - 
\hbar {\partial f \over \partial x'} \Big) =  
-{\hbar^2 \over 2m} {\partial^2 \psi' \over \partial x'^2} +   
 {\hbar^2 \over 2m} \Big( {\partial f \over \partial x'}\Big)^2 
   -  {i \hbar^2 \over m} {\partial f \over \partial x'}{\partial \psi' 
\over \partial x'} - {i \hbar^2 \over 2m} {\partial^2 f \over \partial x'}
\psi' \:.  \label{GENERIC}
\end{equation}  
If the function $f(z't')$ is now choosen of the form
\begin{equation}
f(z',t')=- {mv \over \hbar} \bigg(z'+ \frac{vt'}{2}\bigg) - 
           {m a t' \over \hbar} \bigg( z' +\frac{vt'}{2}+\frac{at'^2}{6}
\bigg) \:, \label{PHASE}
\end{equation}
Eq. (\ref{GENERIC}) simplifies as follows:
\begin{equation} 
i \hbar {\partial \psi' \over \partial t' }(z',t') = - {\hbar^2 \over 2m} 
{\partial^2 \psi' \over \partial {z'}^2} (z',t') + m a z' \psi'(z',t') \:. 
\label{SCHROD2}
\end{equation}
The second term in the righthandside of Eq. (\ref{SCHROD2}) represents 
the effect of an effective potential $V_{in}(z')= m a z'$ which, in the 
classical limit, corresponds to the well known inertial force for the 
case of constant acceleration. 
By putting $a=0$, we recover the Galileo transformation between two inertial 
frames of reference translating with relative velocity $v$. In this case 
the invariance of Eq. (\ref{SCHROD}) reflects, as expected, the 
validity of the Galileian relativity principle \cite{LANDAU}. According 
to (\ref{ANSATZ}), the wavefunctions are related in this case by the 
transformation
\begin{equation}
\psi'(z',t')=\exp\bigg\{ - {i \over \hbar} \Big(mvz' +{v^2 t'\over 
2}\Big) \bigg\} \psi(z'+vt',t') \:.  \label{GALI}
\end{equation} 
One can easily check that the phase factor involved in (\ref{GALI}) is 
just the one needed to ensure the correct transformations of the average 
values of position and momentum, namely
\begin{equation} 
\langle \hat{z}' \rangle=\langle \hat{z} \rangle - vt \:, \hspace{1cm}
\langle \hat{p}' \rangle=\langle \hat{p} \rangle - mv \:,
\end{equation}
in agreement with the Heisenberg equations of motion:
\begin{equation}
{d \hat{z}' \over dt } = {d \hat{z} \over dt}-v  = {\hat{p}' \over m} \:, 
\hspace{1cm}
{d \hat{p}' \over dt } = {d \hat{p} \over dt } = 0 \:. 
\end{equation}
Finally, we write the density matrix associated to the pure state 
(\ref{ANSATZ}) with $f$ given by (\ref{PHASE}) as
\begin{equation}
\rho'(z_1',z_2',t')=\exp\bigg\{ -{i \over \hbar}\Big( m v + ma t' \Big) 
(z_1'-z_2') \bigg \}\,   
\rho\Big(z_1'+v t'+\frac{1}{2} a t'^2,z_2'+v t'+\frac{1}{2} a 
t'^2,t'\Big)  \:.
\end{equation}


\mediumtext
\begin{table}
\caption{
Predicted standard deviations for the times of flight (in msec) of freely 
falling Gaussian states and Schr\"odinger cat (male, $c_+$=$c_-$=1 and 
female, $c_+$=$-c_-$=1) states, starting at rest from an height 
$z_0=3$ mm. The corresponding average time of flight is $T_{of}=24.74$ ms and 
the values $\Delta=\Delta_0=100$ nm have been chosen.} 
\begin{tabular}{l|c|c|c} 
              & Gaussian state & Male cat state & Female cat state 
  \\ \hline
  He$\mbox{}^*$\tablenote{$\,\mbox{}^3$S metastable Helium.} 
              & 10.86   &  7.38 & 15.98 \\
     Be       &  4.82   &  3.28 &  7.09 \\
     Na       &  1.89   &  1.29 &  2.78 \\
     Rb       &  0.36   &  0.24 &  0.53 \\
     Cs       &  0.33   &  0.22 &  0.48  
\end{tabular} 
\end{table}

\end{document}